\documentclass[oribibl]{llncs}
\usepackage{dsfont}
\usepackage{amssymb}
\usepackage{amstext}
\usepackage{graphicx}
\usepackage{algorithm}
\usepackage{algorithmic}

\usepackage[english]{babel}

\input xy
\xyoption{all}

\usepackage{algorithm}
\usepackage{algorithmic}

 \newcommand{\C}{\mathcal{C}}
 \newcommand{\G}{\mathcal{G}}
 \newcommand{\D}{\mathcal{D}}

 \newcommand{\HH}{\mathcal{H}}

 \newcommand{\comp}{\circ}

\begin{document}

\title{Multi-diagrams of relations between fuzzy sets: weighted limits, colimits and commutativity}

\author{Carlos Leandro\inst{1} \and Lu\'{\i}s Monteiro\inst{2}}

\institute{Departamento de Matem\'{a}tica, Instituto Superior de Engenharia de Lisboa, Portugal.
\and CITI, Departamento de Inform\'{a}tica,  Faculdade de Ci\^{e}ncias e Tecnologia, \\Universidade Nova de Lisboa, 2829-516 Caparica, Portugal.
}

\maketitle

\begin{abstract}
Limits and colimits of diagrams, defined by maps between sets, are universal constructions fundamental in different mathematical domains and key concepts in theoretical computer science. Its importance in semantic modeling is described by M. Makkai and R. Par\'{e} in \cite{Makki89}, where it is formally shown that every axiomatizable theory in classical infinitary logic can be specified using diagrams defined by maps between sets, and its models are structures characterized by the commutativity, limit and colimit of those diagrams. Z. Diskin in \cite{Dis97}, taking a more practical perspective, presented an algebraic graphic-based framework for data modeling and database design. The aim of our work is to study the possibility of extending these algebraic frameworks to the specification of fuzzy structures and to the description of fuzzy patterns on data.  For that purpose, in this paper we describe a  conservative extension for the notions of diagram commutativity, limit and colimit, when diagrams are constructed using relations between fuzzy sets, evaluated in a multi-valued logic. These are used to formalize what we mean by ``a relation $R$ is similar to a limit of diagram $D$,'' ``a similarity relation $S$ is identical to a colimit of diagram $D$ colimit,'' and ``a diagram $D$ is almost commutative.''
\end{abstract}

%----------------------------------------------------------------------
\section{\uppercase{Introduction}}

The most general universe of current mathematical discourse is the category known as $Set$, whose objects are sets and whose arrows are the set functions\cite{Goldblatt06}. It is the universe by default for the construction of models for mathematical theories. Here the fundamental mathematical concepts such as number and relation are given formal descriptions, and the specification of axioms legislating about  the properties of sets leads to a so called foundation of mathematics. The basic set-theoretic operations and attributes such as empty set, intersection, product set and surjective function, can be described by reference to the arrows in $Set$, and these descriptions have been interpreted in any Category.

Set-theoretic notions are governed by the classic boolean logic, and have been, with success, applied to the description of some human activities, governed by this logic, like data specification and database design. Z. Diskin in \cite{Dis97}, formalized these types of applications, by presenting an algebraic graphic-based framework for data modeling and database design.  This type of methodology was initially explored by Ehresmann\cite{Ehresm68} for algebraic specification. He developed a structure, named a \emph{sketch}, as an alternative to the string-based specification employed in mathematical logic. A \emph{sketch}, in Ehresmann sense, is defined by a category and three sets of diagrams defined in the category: a set of commutative diagrams, a set of limit cones and a set of colimit cocones. Defining a model for the sketch by a functor, in $Set$, preserving the sketch structure, i.e. commutativity, limits and colimits,   Makkai and R. Par\'{e}, in \cite{Makki89}, showed that this structure is absolutely expressive. Every mathematical theory can be formalized using a sketch, and the category of its models can be presented as a category of functors preserving the sketch structure \cite{Adamek94}.

Modern human activities impose the description of structures similar to set-theoretic notions, but that are not governed by classical logic. This, in some sense, explains the increasing importance of probabilistic models in our daily life. These models can be seen as patterns that are present on data, and its use is usually governed by probabilistic logic or a fuzzy logic.  We centered our work on the description of fuzzy structures. And given the descriptive power of algebraic tools like sketches, we work on the possibility of performing this description using limits, colimits and commutativity.  For it, this paper presents fuzzy conservative extension to these set-theoretic notions. These notions are described in a general universe for fuzzy modeling given by a class of structures, denoted by $Rel_\Omega$, having by morphisms relations evaluated in a   multi-valued logic $\Omega$, and where composition is defined using a semiring defined using logic connectives. Object in this category are defined by a membership relation and a similarity relation, encoding the degree of truth for ``$x\in X$'' and describing the degree of truth for proposition ``$x=y$.'' In $Rel_\Omega$ morphism are conservative bimodules, a type of relation evaluated in $\Omega$, which conserves membership and similarity degrees between target and source objects.

The universe $Set$ is a substructure of $Rel_\Omega$. In the following we present conservative extensions in $Rel_\Omega$ to the notions of limit, colimit and commutativity in $Set$, in the sense that when a diagram is defined using maps between classic sets, the described extensions coincide with the categorical ones. Furthermore,  our approach allows the extension of Ehresmann's sketch structure in two directions. We propose a logic extension, used to specify propositions like ``a relation $R$ is $\lambda$-similar to diagram $D$ limit,'' ``the similarity relation $S$ is $\lambda$-identical to a diagram $D$ colimit,'' and by a ``diagram $D$ is $\lambda$-almost commutative.'' We also propose a functional extension, where instead of diagrams we use multi-diagrams for the  graphic-based proposition description. Here we assume a multi-graph as a structure defined by arrows linking two sets of vertices. The use of this graphic representation to construction of models impose the existence of a rich interpretation framework, named of multi-category. Categories like $Rel_\Omega$ have this structure characterized by the existence of operators for the construct complex objects and morphisms from simplest ones, and the existence of an operator on arrows, defined everywhere having by restriction the composition operator. In this context we see a multi-diagram as a circuit defined by aggregation of multi-morphisms.

The rapid development of computer technology in the last decades has made
it possible to easily collect huge amounts of data. Analyzing such
large data sets is tedious and costly, and thus we need eficiente methods to
be able to understand how the data was generated, and what sort of patterns
or regularities exist in the data.
In order to find patterns or regularities in the data, it is necessary
that we can describe how far from each other two data objects are. This
is the reason why similarity between objects is one of the central concepts
in knowledge discovery. During the last few years,
there has been considerable interest in defining intuitive and easily computable
measures of similarity between objects in diference application areas
and in using abstract similarity notions in querying databases.
This becomes one of the major areas of research in database modeling. There has been the continuous effort to enrich existing data base models with a more extensive collection of
semantic concepts. One of the semantic needs not adequately addressed by
traditional models is that of imprecision and uncertainty.  Traditional models
assume the database model to be a correct reflection of the world being
captured and assume that the data stored is known, accurate, and complete. It
is rarely the case in real life that all or most of these assumptions are met.
Different data models have been proposed to handle different categories of
data quality (or lack thereof) and query evaluation with fuzzy set theory.
In our approach to data modeling and database design was centered in the semantic extension of Ehresmann sketches described on the category $Rel_\Omega$, where objects are characterized by a membership relation and a similarity relation both evaluated in the same multi-valued logic.

\section{\uppercase{Preliminaries}}
We begin by presenting the basic notions needed on the definition of multi-diagram and multi-category.

\subsection{Multi-categories}
A circuit defines a relation between its input values (the values carried
on its source wires) and its output values (the values carried on its target wires). Any two points on the same wire are constrained to carry equal values, while components impose more complex constraints on the values carried by their input and output wires. We view the discrete components of a circuit as morphism in a category whose objects are set of types and build complex circuits from the basic components using a gluing operation. The notion of multi-category tries to capture the possibility of morphisms and objects in a category to be generated by aggregation of simplest morphisms and objects. Its name was inspired by the notion of multi-limit propose by Dires' on the context of free product completion. This is a property common to many categories making them adequate as framework for circuit modeling. Example of this are the locally presentable categories where objects can be generated using limit having by vertices representable objects \cite{Adamek94}. This structure can emerge naturally associated to structural completion of some categories, the following examples describe this idea.

Let $\C$  be a category (see definition in \cite{Borceux94}), having $\C$-objects in $\C_0$ and arrows or $\C$-morphisms in $\C_1$. If $f:A\rightarrow B$ is a morphism in $\C_1$ we write $f\in \C[A,B]$.  For each object $A$ in $\C_0$ its identity is in $\C_1$ and will be denoted by $1_A:A\rightarrow A$.

We used as a model for our notion of multi-category the structure generated from the Diers' product completion process.

\begin{example}[Diers' completion]
Given a category $\C$ with objects in $\C_0$ and morphism in $\C_1$. For all morphism $f\in \C_1$ let $dom(f)$ and $codom(f)$ be respectively $f$ domain and codomain. Given two morphisms $g:A\rightarrow B$ and $f:B\rightarrow C$, its composition is a morphism $f\circ g:A\rightarrow C$.  We denote by $set(\C)$ the structure:
\begin{enumerate}
  \item having by objects sets of $\C$-objects $\bar{A}\subset \C_0$, the class of its objects is denoted by $set(\C_0)$,
  \item having by arrows $\bar{f}:\bar{A}\rightarrow \bar{B}$ sets of morphisms $\bar{f}\subset \C_1$ such that for every $f\in \bar{f}$, $dom(f)\in \bar{A}$ and $codom(f)\in \bar{B}$, the class of its arrows is denoted by $set(\C_1)$,
  \item arrows in $set(\C)$ can be operated extending composition in $\C$, given $\bar{f}:\bar{A}\rightarrow \bar{B}$ and $\bar{g}:\bar{C}\rightarrow \bar{D}$, we have $\bar{h}=\bar{f}\circ \bar{g}:\bar{E}\rightarrow \bar{F}$, where $\bar{E}=\bar{C}\cup\bar{A}\setminus\bar{D}$ and $\bar{F}=\bar{B}\cup\bar{D}\setminus\bar{A}$, such that
\[
\bar{f}\circ\bar{g}=
\begin{array}{l}
  \{g\in\bar{g}: codom(g)\notin\bar{A}\}\cup \\
  \cup\{f\in\bar{f}: dom(f)\notin\bar{A}\}\cup \\
  \cup \{f\circ g: f\in\bar{f}, g\in\bar{g}, dom(f)=codom(g)\}
\end{array}
\]
  \item In $set(\C)$ we can identify unary operators
  \begin{enumerate}
    \item $id:set(\C_0)\rightarrow set(\C_1)$, assigning to each set of objects $\bar{A}$ a set $\bar{f}:\bar{A}\rightarrow\bar{A}$ of identity morphisms, such that for every $A\in\bar{A}$, $1_A\in \bar{f}$
    \item $\_\Box:set(\C_1)\rightarrow set(\C_0)$ and $\Box\_:set(\C_1)\rightarrow set(\C_0)$, defined by $\Box\bar{f}=\bar{A}$ and $\bar{f}\Box=\bar{B}$ if $\bar{f}:\bar{A}\rightarrow\bar{B}$
  \end{enumerate}
  \item We can identify constants
  \begin{enumerate}
    \item $\top\in set(\C_0)$, defined by the empty set of objects and
    \item $1_\top\in set(\C_1)$ the empty set of morphisms
  \end{enumerate}
\end{enumerate}
In the structure $set(\C)$ we have:
\begin{enumerate}
  \item since $set(\C_0)$ is defined using sets of objects the relation $\subset$ defined a partial order;
  \item when we restrict the operator $\circ$ to componible arrows we define a category having by objects sets of objects $\C_0$ and by morphisms $\C_2$. Given an object $\bar{A}\in set(\C_0)$, $id(\bar{A})=1_{\bar{A}}\in set(\C_1)$ is a set of identity morphisms, and for every morphism $\bar{f}\in set(\C_1)$, $dom(\bar{f})=\Box \bar{f}$ and $codom(\bar{f})= \bar{f} \Box$;
  \item Naturally we have $1_\top=id(\top)$;
  \item For every $\bar{A}\in \C_0$, $\Box id(\bar{A})= id(\bar{A})\Box = \bar{A}$;
  \item The operator $\circ$ induces a monoidal structure in $\C_2$, having by identity $1_\bot$ and for every pair of multi-morphisms $\bar{f},\bar{g}\in \C_2$ is valid
            \begin{equation}
                \Box(\bar{f}\circ \bar{g})=\Box \bar{g} \cup \Box \bar{f }\backslash \bar{g}\Box,\text{ and }
                (\bar{f}\circ \bar{g})\Box=\bar{f}\Box \cup \bar{g}\Box \backslash \Box \bar{f}.
            \end{equation}
\end{enumerate}

For instance, for distinct objects $A$, $B$, $C$ and $D$, in $\C$, we have in $set(\C)$ $\{f:A\rightarrow B\}\circ \{f:A\rightarrow B\} = \{f:A\rightarrow B\}$,$\{f:B\rightarrow C\}\circ \{g:A\rightarrow B\} = \{f\circ g:A\rightarrow C\}$, $\{f:C\rightarrow D\}\circ \{g:A\rightarrow B\} = \{f:C\rightarrow D, g:A\rightarrow B\}$ and $\{f:B\rightarrow C,h:C\rightarrow D\}\circ \{g:A\rightarrow B\} = \{h:C\rightarrow D, f\circ g:A\rightarrow C\}$.

We denote by $J_\C:\C\rightarrow set(\C)$ the canonical embedding which identifies objects of $\C$ with singleton sets.
\end{example}

For every category $\C$, we called to structure $set(\C)$ the $\C$ \emph{free multi-category completion}. However instead of making the completion using sets of morphisms, in the original description of product completion by Diers in \cite{Diers77}  he used families of morphisms:

\begin{example}[Free product completion]
The free product completion $\prod(\C)$ of a category $\C$ is a structure having
\begin{enumerate}
  \item $fam(\C)$ the class of objects $\bar{A}=(A_i)_{i\in I}$ given by small-indexed families of $\C$-objects $A_i$;
  \item a morphism $\bar{f}:\bar{A}\rightarrow \bar{B}$ in $\prod(\C)$ is given by a function $\varphi:J\rightarrow I$, if $\bar{A}=(A_i)_{i\in I}$ and $\bar{B}=(B_j)_{j\in J}$, and by a small-indexed families of $\C$-morphisms $f_j:A_{\varphi(j)}\rightarrow B_j$ $(j\in J)$. The class of arrows in $\C$ is denoted by $fam(\C)$;
  \item given $\bar{f}=(f_j:A_{\varphi(j)}\rightarrow A_j)_J$, described by $\varphi:J\rightarrow I$, and $\bar{g}=(g_j:A_{\psi(j)}\rightarrow A_j)_J$, described by $\psi:I\rightarrow N$, a possible extension for composition in $\C$ is given defining $\bar{h}=\bar{g}\circ \bar{f}$, where $\bar{h}=(h_t:A_{\alpha(t)}\rightarrow A_t)_T$, with $\alpha:J\cup I\setminus L\rightarrow N\cup L\setminus I$ such that:
        \begin{enumerate}
            \item $\alpha(j)=\varphi(j)$ and $h_j=f_j$, if $\varphi(j)\in L\setminus I$,
            \item $\alpha(j)=(\psi\circ\varphi)(j)$ and $h_j=g_{\varphi(j)}\circ f_j$, if $\varphi(j)\in I\cap L$, and
            \item $\alpha(j)=\psi(j)$ and $h_j=g_j$, if $j \in I\setminus L$.
        \end{enumerate}

  \item In the structure $\prod(\C)$ we may identify unary operators:
  \begin{enumerate}
    \item $id:fam(\C_0)\rightarrow fam(\C_1)$ assigning to each family of objects $\bar{A}=(A_i)_I$ a family $\bar{f}=(1_i:A_{1_I(i)}\rightarrow A_i)$ of identity morphisms, described by the identity $1_I:I\rightarrow I$;
    \item $\_\Box:fam(\C_1)\rightarrow fam(\C_0)$ and $\Box\_:fam(\C_1)\rightarrow fam(\C_0)$, defined by $\Box\bar{f}=(A_i)_I$ and $\bar{f}\Box=(B_j)_J$, if $\bar{f}=(f:A_{\varphi(j)}\rightarrow B_j)_J$ described by $\varphi:J\rightarrow I$.
  \end{enumerate}
  \item In $\prod(\C)$ we have by constantes:
  \begin{enumerate}
    \item $\top\in fam(\C_0)$, defined by the empty family of objects and
    \item $1_\top\in fam(\C_1)$ the empty family of morphisms.
  \end{enumerate}
\end{enumerate}
On the structure $\prod(\C)$
\begin{enumerate}
  \item $fam(\C_0)$ is partially ordered by a relation $\subset$, such that $(A_i)_I\subset (B_j)_J$ if for every $i\in I$ exists $j\in J$ such that $A_i=B_j$. On the class of families of $\C$-objects, we define $(A_i)_I\cup (A_j)_J= (A_k)_{k\in I\coprod J}$  and $(A_i)_I\setminus (A_j)_J= (A_k)_K$ if $K\subset I$ is such that $i\in K$ iff for every $j\in J$, $A_j\neq A_i$.
  \item $1_\top=id(\top)$ and $i_0(\top)=\top$;
  \item For every $(A_i)_I\in fam(\C_0)$, $\Box id((A_i)_I)= id((A_i)_I)\Box = (A_i)_I$;
  \item The operator $\circ$ induces a monoidal structure in $fam(\C_2)$, having by identity $1_\bot$ and for every pair of arrows $\bar{f},\bar{g}\in fam(\C_2)$ is valid
            \begin{equation}
                \Box(\bar{f}\circ \bar{g})=\Box \bar{g} \cup \Box \bar{f }\backslash \bar{g}\Box,\text{ and }
                (\bar{f}\circ \bar{g})\Box=\bar{f}\Box \cup \bar{g}\Box \backslash \Box \bar{f}.
            \end{equation}
\end{enumerate}

For every pair of objects $\bar{A}=(A_i)_{i\in I}$ and $\bar{B}=(B_j)_{j\in J}$, its product in $\prod(\C)$ is given by $\bar{A}\times \bar{B}=(C_l)_{l\in I\amalg J}$, where $C_l=A_l$ if $l\in I$ and $C_l=B_l$ if $l\in J$, and having projections defined by families of identities and described using, respectively, coprojections $p_1:I\rightarrow I\amalg J$ and $p_2:J\rightarrow I\amalg J$.

When $\C$ has products, there is a product-preserving functor $\Pi:\;\prod(\C)\rightarrow \C$ assigning to each object $\bar{A}$ its product $\prod\bar{A}$ in $\C$. This functor describes a isomorphism between category $\C$ and $\prod(\C)$. In this case we use the product to aggregate families of objects and families of arrows. This is done using two maps $i_0:\C_0\rightarrow \C_0$ and $i_2:\C_2\rightarrow \C_2$ defined by $i_0((A_i)_I)=\prod_{i\in I}A_i$ and If when $\bar{f}=(f_j)_J:\Box \bar{f} \rightarrow \bar{f}\Box$ in $\C_2$, we define $i_2(\bar{f})=\prod_{j\in J} f_j:i_1(\Box f)\rightarrow i_1(f\Box)\in \C_1$.
Note that $i_1$ and $i_2$ are idempotents, $i_0\circ i_0 = i_0$ and $i_2\circ i_2 = i_2$, and both maps have a nonempty class of fixed points. Every $\C$-object is a fixed point for $i_1$ and every $\C$-morphism is a fixed point for $i_2$. And in particular, we have $i_0(\top)$ is a $\C$ terminal object and $i_0(1_\top)$ is its identity.
\end{example}

Bellow we present another example of a category with multi-category structure, in this case inspired on the free generation of circuits based on a set of designated componentes.

\begin{example}[Library]
Let $\Sigma$ be a set of symbols, and let $\Sigma^+$ be the associated \emph{polarized alphabet}. The set of words generated by the polarized alphabet $\Sigma^+$ will be denoted by $(\Sigma^+)^\ast$.

In $(\Sigma^+)^\ast$ the \emph{gluing} of word $w$ and $w'$ is the word $w\otimes w'$ what results from applying Algorithm \ref{gluing}.

\begin{algorithm}
\caption{Gluing words} \label{gluing}
\begin{algorithmic}
\STATE 1. \textbf{Input}: $w,w'\in(\Sigma^+)^\ast$
\STATE 2.\textbf{Output}: $w\otimes w'$
\STATE 3. Let $w_0=w$, $w'_0=w'$ and $i=0$;
\STATE 4. Let $\lambda$ be the first output symbol in $w_i$ with its dual $\lambda^+$ in $w'_i$;\label{ckpoint}
\STATE 5. Generation of $w_{i+1}$ by deleting the first occurrence of $\lambda$ in $w_i$;
\STATE 6. Generation of $w'_{i+1}$ by deleting the first occurrence of $\lambda^+$ in $w'_i$;
\STATE $i=i+1$;
\STATE 7. Repeat \ref{ckpoint}, while exits a symbol in $w_i$ having its dual in $w'_i$;
\STATE 8. $w\otimes w'=w_iw'_i$ by word concatenation;
\end{algorithmic}
\end{algorithm}

The set $\Sigma^+$ can be factorized as $\Sigma^+=\Sigma_O\cup \Sigma_I$, such that $s\in \Sigma_O$ iff $s^+\in \Sigma_I$. Defining for every word $w\in (\Sigma^+)^\ast$, $i(w)=\{s\in \Sigma_I:\text{ symbol } s \text{ is used on }w\}$ and $o(w)=\{s\in \Sigma_I:\text{ symbol } s^+ \text{ is used on }w\}$, we have
$i(w\otimes w')=i(w) \cup i(w') \backslash o(w)$ and  $o(w\otimes w')=o(w') \cup o(w) \backslash i(w')$.

A \emph{library of componentes} $L$ is a map from a set $I$ to $(\Sigma^+)^\ast$, $L:I\rightarrow (\Sigma^+)^\ast$. Every $l\in I$, defines a \emph{componente} in $L$ having its signature codified in a word $L(l)$. The selection of a componente $l\in I$ defines a multi-morphism $\bar{f}$ having by source the set $\Box\bar{f}=i(L(l))$ and by target $\bar{f}\Box$ is the set of symbols dual to output symbols, $\bar{f}\Box=o(L(l))^+$. In this case we denoted a multi-morphism using a triple $\bar{f}=(l,\Box\bar{f},\bar{f}\Box)$.

Given a library $L:I\rightarrow (\Sigma^+)^\ast$, we define its free monoidal completion as a library $L^\ast:I^\ast\rightarrow (\Sigma^+)^\ast$, where $I^\ast$ is the set of sequences $\langle c_1,c_2,\ldots,c_n\rangle$ with components $c_i$ in $I$, such that:
\begin{enumerate}
  \item $L^\ast(l)=L(l)$ if $l\in I$,
  \item $L^\ast(ww')=L^\ast(w)\otimes L^\ast(w')$ if $w,w'\in I^\ast$ and
  \item $L^\ast(\top)=\top$ (the empty word have an empty signature).
\end{enumerate}
We can see $L^\ast:I^\ast\rightarrow (\Sigma^+)^\ast$ as a library generated from componentes on $L$, in this sense we named it of \emph{library of circuits}.

Let $L^\ast:I^\ast\rightarrow (\Sigma^+)^\ast$ be a library of circuits with components in the library $L:I\rightarrow (\Sigma^+)^\ast$. The correspondence $L^*$ has associated a structure described by:
\begin{enumerate}
  \item a class of objects $\C_0$ defined by sets of input symbols $\Sigma$,
  \item a class of arrows $\C_2$ defined by sequences $\langle c_1,c_2,\ldots,c_n\rangle$ of componentes in $I^*$,
  \item a binary operator defined using concatenation
  \[\circ:\C_2\times\C_2\rightarrow \C_2,
  \]
  such that $c_1\circ c_2=c_1c_2$;
  \item unary operators
  \begin{enumerate}
    \item $id:\C_0\rightarrow \C_2$ such that for every set on symbols $A=\{s_0s_1\ldots s_n\}$ we define $id(A)=1_A$ a diagram such that $L^*(1_A)=s_0s_1\ldots s_ns_0^+s_1^+\ldots s_n^+$,
    \item For every $c_1\in \C_2$, we define $c_1\Box=o(L^*(c_1))$ and $\Box c_1=o(L^*(c_1))$
  \end{enumerate}
  \item The empty set defines an object $\top\in\C_0$ and is identity is denoted $1_\top\in\C_2$
\end{enumerate}
This structure satisfies:
\begin{enumerate}
  \item $\C_0$ is partially ordered by set inclusion, and has monoidal structure defined by set union;
  \item The class $\C_2$ and operator $\circ$ define a category having by objects $\C_0$, by morphisms $\C_2$, and by composition the restriction of $\circ$ to composable morphisms. Given an object $A\in \C_0$, $id(A)=1_A\in\C_2$ is the identity morphism, and for every morphism $c_1\in \C_2$, $codom(c_1)= c_1 \Box$ and $dom(c_1)=\Box c_1$;
  \item $1_\top=id(\top)$ and $i_0(\top)=\top$;
  \item For every $A\in \C_0$, $\Box id(A)= id(A)\Box = A$;
  \item The operator $\circ$ induces a monoidal structure in $\C_2$, having by identity $1_\top$ and for every pair of multi-morphisms $c_1,c_2\in \C_2$ is valid
            \begin{equation}
                \Box(c_1\circ c_2)=\Box c_2 \cup \Box c_1 \backslash c_2\Box,\text{ and }
                (c_1\circ c_2)\Box=c_1\Box \cup c_2\Box \backslash \Box c_1.
            \end{equation}
\end{enumerate}

A semantic for a library $L:I\rightarrow (\Sigma^+)^\ast$ in a category with products $\D$, is defined assigning symbols to objects, by a map $\gamma_1:\Sigma\rightarrow \D_0$ and components to arrows, using a map $\gamma_2:I\rightarrow \D_1$. This assignments must satisfy the consistence principle:

If $c_1:\Box c_1 \rightarrow c_1\Box$ in $\C_2$, then $\gamma_2(c_1):\gamma_1(\Box c_1)\rightarrow \gamma_1(c_1\Box)\in \D_2$.

This interpretation can be extended to a library of circuits with components in $L$, and this extension is unique up to natural isomorphism and can be defined inductively by
\begin{enumerate}
  \item $\gamma_2^\ast(l)=\gamma_2(l)$ if $l\in I$,
  \item $\gamma_2^\ast(ww')=\gamma_2^\ast(w)\circ \gamma_2^\ast(w')$ if $w,w'\in I^\ast$ and
  \item $\gamma_2^\ast(\top)=\top$ (the empty word have an empty signature).
\end{enumerate}
The class of sets, defined by all circuits with the same interpretation, $[c_1]=\{c_2:\;\gamma_2(c_1)=\gamma_2(c_2)\}$ is a partition of $C_2$,  defining an equivalence relation between circuits. For this partition we assume the existence of a choice function $i_2:C_2\rightarrow C_2$ such that $i_2(c_1)=c_2$ if $c_2\in [c_1]$ and for every $c_3\in [c_1]$, $i_2(c_3)=c_2$. Note what, if $f:\Box f \rightarrow f\Box$ in $\C_2$, then $i_2(f):\Box f\rightarrow f\Box\in \C_2$.

When $\D$ has products for every $f:\Box f \rightarrow f\Box$ in $\C_2$, we have $i_2(f):\prod(\Box f)\rightarrow \prod(f\Box)\in \C_2$.
\end{example}

We generalized the structure presented on this examples defining the notion of multi-category.

\begin{definition}[Multi-category]
A multi-category $\D$ is defined by
\begin{enumerate}
  \item a class of objects $\D_0$ (the class of objects or circuits),
  \item a class of multi-morphisms $\D_2$ (the class of multi-morphisms or links),
  \item a binary operator
  \[\circ:\D_2\times\D_2\rightarrow \D_2\text{ the gluing operator},
  \]
  \item unary operators
  \begin{enumerate}
    \item $i_0:\D_0\rightarrow \D_0$ (object aggregation),
    \item $i_2:\D_2\rightarrow \D_2$ (multi-morphism aggregation),
    \item $id:\D_0\rightarrow \D_2$ (the identity),
    \item $\_\Box:\D_2\rightarrow \D_0$ (sources), and
    \item $\Box\_:\D_2\rightarrow \D_0$ (targets)
  \end{enumerate}
  \item constantes
  \begin{enumerate}
    \item $\top\in\D_0$, (empty circuit) and
    \item $1_\top\in\D_2$ (empty wire)
  \end{enumerate}
\end{enumerate}
such that
\begin{enumerate}
  \item $\D_0$ is a partially ordered set by a relation $\subset$, where is defined a monotone operator $\cup$ and such that
      \begin{enumerate}
        \item $(\D_0,\cup, \top)$ is a monoid and
        \item there is a binary operator $\setminus$ defined in $\D_0$ such that:
            \[
            B\setminus A\subset C \Leftrightarrow B\subset A\cup C ,
            \]
            for every $A,B,C\in \D_0$.
      \end{enumerate}
  \item the operators $i_0$ and $i_1$ are idempotents, i.e. $i_0\circ i_0 = i_0$ and $i_2\circ i_2 = i_2$;
  \item $i_2$ have a nonempty class $\D_1$ of fixed points, for $\D_1=\{f:i_2(f)=f\}\subseteq \D_2$. Multi-morphisms in $\D_1$ are named morphisms;
  \item The class $\D_1$ and operator $\circ$ define a category having by objects $\D_0$, by morphisms $\D_1$, and by composition the restriction of $\circ$ to composable morphisms. Given an object $A\in \D_0$, $id(A)=1_A\in\D_1$ is the identity morphism, and for every morphism $f\in \D_1$, $dom(f)=\Box f$ and $codom(f)= f \Box$;
  \item $1_\top=id(\top)$ and $i_0(\top)=\top$;
  \item For every $A\in \D_0$, $\Box id(A)= id(A)\Box = A$;
  \item If $f:\Box f \rightarrow f\Box$ in $\D_2$, then $i_2(f):i_0(\Box f)\rightarrow i_0(f\Box)\in \D_1$;
  \item The operator $\circ$ induces a monoidal structure in $\D_2$, having by identity $1_\top$ and for every pair of multi-morphisms $f,g\in \D_2$ is valid
            \begin{equation}
                \Box(f\circ g)=\Box g \cup \Box f \backslash g\Box,\text{ and }
                (f\circ g)\Box=f\Box \cup g\Box \backslash \Box f.
            \end{equation}
\end{enumerate}
\end{definition}

%As usual in the class $\C_0$ we define for every par of objects $A$ and $B$, $A\cap B=(A\cup B)\setminus ((A\setminus B)\cup (B\setminus A))$.

Naturally, every multi-category can be seen as a category, having the some objects and by morphisms the multi-morphism on the multi-category.

A multi-functor between multi-categories $\D$ and $\HH$, is denoted by $F:\D\rightarrow \HH$, and is defined using a pair of maps $(F_0,F_2)$ such that $F_0:\D_0\rightarrow \HH_0$ is a transformation between objects and $F_2:\D_2\rightarrow \HH_2$ is a transformation between morphisms such that
\begin{enumerate}
  \item $F_0(i_0(A))=i_0(F_0(A))$, for every $A\in \D_0$;
  \item $F_2(i_2(f))=i_2(F_2(f))$, for every $f\in \D_2$;
  \item $F_2(f\comp g)=F_2(f)\comp F_2(g)$;
  \item if $f:A\rightarrow B$ then $F_2(f):F_0(A)\rightarrow F_0(B)$;
  \item $F_0(\top)=\top$;
  \item $F_0(A\cup B)=F_0(A)\cup F_0(B)$;
  \item $F_2(1_A)=1_{F_0(A)}$
\end{enumerate}

A multi-category can be seen as a way to extend the structure of a category. We interpret a multi-category as the structural extension for the category defined by fixed points to the object and morphism aggregation maps. In this sense we define:

\begin{definition}
Let $\C$ be a category and $J:\C\rightarrow \D$ an embedding where $\D$ is a multi-category. The multi-category $\D$ is generated by $J(\C)$ if:
 \begin{enumerate}
   \item the object aggregation map $i_0$ has by fixed points $J(\C_0)$ the class of images of $\C$-object, and
   \item the morphism aggregation map $i_2$ has by fixed points $J(\C_1)$ the class of images of $\C$-morphisms.
 \end{enumerate}
\end{definition}

Since every multi-category have structure of category every functor $F:\C\rightarrow \HH$ defined from a category $\C$ to a multi-category $\HH$, can be extended to a multi-functor between a multi-category $\overline{F}:\D\rightarrow \HH$, where $\D$ must be a multi-category generated by $J(\C)$. This extension is unique, up to isomorphism, and is defined by $\overline{F}(A)=A$ and $\overline{F}(f)=f$ if $A$ and $f$ are objects and morphisms in $J(\C)$, $\overline{F}(A\cup B)=\overline{F}(A)\cup \overline{F}(B)$, $\overline{F}(\top)=\top$ and $\overline{F}(f\comp g)=\overline{F}(f)\comp \overline{F}(g)$. In this sense we have:

\begin{proposition}
Given multi-categories $\D$ and $\HH$, where $\D$ is generated by $J(\C)$, for every functor $F:\C\rightarrow \HH$ there is a unique up to isomorphism multi-functor $\overline{F}:\D\rightarrow \HH$ such that $$F=\overline{F}\circ J.$$
\end{proposition}

We see a multi-category as a completion for a category when a structure of multi-category can be identify in the category.

\begin{definition}
A category $\C$ has the structure of multi-category if there is a multi-category $\D$ such that there is an isomorphism $\C$ and $\D$, defined by a functor $J:\C\rightarrow\D$.
\end{definition}

By this we mean what, there is an embedding $J:\C\rightarrow\D$ such that every $\D$-object $D$ there is a $\C$-object $C$ such that $J(C)\cong D$.

\subsection{Monoidal symmetric category}
A category $\C$ is \emph{monoidal symmetric}, if there is a bifunctor $-\otimes -:\C\times\C\rightarrow\C$, the \emph{tensor product}, an object $\top$, the \emph{unit}, and natural isomorphisms $\alpha_{XYZ}:(X\otimes Y)\otimes Z\rightarrow X\otimes (Y\otimes Z)$, $r_X:X\otimes\top\rightarrow X$, $l_X:\top\otimes X\rightarrow X$, and $s_{XY}:X\otimes Y\rightarrow Y\otimes X$, satisfying some coherence conditions (see the bifunctor lemma in \cite{Awodey94}).

\begin{definition}
A complete residuated lattice (CRlattice for short) is an algebra $\mathbf{\Omega}=(\Omega,\otimes,\Rightarrow,\wedge,\vee,\bot,\top)$ with four binary operations and two constants such that:
\begin{enumerate}
  \item $(\Omega,\wedge,\vee,\bot,\top)$ is a complete lattice with largest element $\top$ and least element $\bot$ (with respect to the lattice ordering $\leq$);
  \item $(\Omega,\otimes,\top)$ is a commutative semigroup with the unit $\top$, i.e. $\otimes$ is commutative, associative and $\top\otimes x = x$ for all $x$;
  \item the residuation equivalence holds:
  \begin{center}
      $z\leq(x\Rightarrow y)$ iff $x\otimes z\leq y$ for all $x,y,z$.
  \end{center}
\end{enumerate}
Since these lattices are complete, for every subset $M\neq \emptyset$ of $\Omega$ we have $\bigvee M\in \Omega$ and $\bigwedge M\in \Omega$.
\end{definition}

CRlattices are basic structures of truth degrees used in fuzzy logic.  A CRlattice is a  BL-algebra if additionally the following conditions hold:   $x\wedge y = x\otimes(x\Rightarrow y)$ (divisibility) and $(x\Rightarrow y)\vee (y \Rightarrow x)= 1$ (pre-linearity). Particularly useful BL-algebras, defined when $\Omega$ is the closed unit real interval, when $x\otimes y=\max(x+y-1,0)$ are used for modeling \emph{{\L}ukasiewicz logic}. \emph{G\"{o}del logic} has as models the BL-algebras described using $x\otimes y=\min(x,y)$, and for product logic it is assumed that $x\otimes y=x.y$ (product of reals), see \cite{Hajek98}.

We can consider a CRlattice  $\mathbf{\Omega}=(\Omega,\otimes,\Rightarrow,\wedge,\vee,\bot,\top)$ as a category, having by objects the elements of $\Omega$, and where arrows are defined by the lattice ordering $\leq$, i.e. there is a unique morphism $\alpha\rightarrow \beta$ iff $\alpha\leq \beta$. We denote its initial and terminal objects by $\bot_\Omega$ and $\top_\Omega$, respectively. Further more, $\mathbf{\Omega}$ is a symmetric monoidal closed category, where the functor $X\otimes-:\mathbf{\Omega}\rightarrow \mathbf{\Omega}$ has right-adjoint $X\Rightarrow -:\mathbf{\Omega}\rightarrow \mathbf{\Omega}$. In this sense every CRlattice defines a category having structure of multi-category.

Note that, the free strict monoidal category generated from a category $\C$, usually denoted by $\sum(\C)$, has structure of multi-category. Its objects are finite sequences $(A_i)_I$ of objects of $\C$. There are arrows between between $(A_i)_I$ and $(B_j)_J$, if and only if, there is a bijection $\alpha:J\rightarrow I$, and then the arrows are families $\bar{f}=(f_j)_J$ such that $f_j:A_{f(j)}\rightarrow B_j$, for every $j\in J$. The tensor product of two objects $(A_i)_I$ and $(A_j)_J$, is given by concatenation  $(A_i)_{I\coprod J}$.  $\bar{f}=(f_j:A_{\varphi(j)}\rightarrow A_j)_J$, described by bijection $\varphi:J\rightarrow I$, and $\bar{g}=(g_j:A_{\psi(j)}\rightarrow A_j)_J$, described by bijection $\psi:I\rightarrow N$, a possible extension for composition in $\C$ is given defining $\bar{h}=\bar{g}\circ \bar{f}$, where $\bar{h}=(h_t:A_{\alpha(t)}\rightarrow A_t)_T$, with $\alpha:J\cup I\setminus L\rightarrow N\cup L\setminus I$ such that:
        \begin{enumerate}
            \item $\alpha(j)=\varphi(j)$ and $h_j=f_j$, if $\varphi(j)\in L\setminus I$,
            \item $\alpha(j)=(\psi\circ\varphi)(j)$ and $h_j=g_{\varphi(j)}\circ f_j$, if $\varphi(j)\in I\cap L$, and
            \item $\alpha(j)=\psi(j)$ and $h_j=g_j$, if $j \in I\setminus L$.
        \end{enumerate}
Note what, in this case $\alpha:J\cup I\setminus L\rightarrow N\cup L\setminus I$ is a bijection, defining $\bar{h}=(h_t:A_{\alpha(t)}\rightarrow A_t)_T$.        

\begin{proposition}
A monoidal symmetric category $(\C,\otimes,\top)$ has the structure of multi-category, defined by the embedding $J:\C\rightarrow \D$, if $$J(A)\cup J(B)\cong J(A\otimes B).$$
\end{proposition}

A monoidal category has structure of multi-category when the tensor product can be seen as the object aggregation in the multi-category.

By a \emph{cartesian category} is meant \cite{Awodey94} a symmetric monoidal category $\C$ having for tensor product $\otimes$ the categorical product $\times$ in $\C$. The monoidal completion was named of free product completion by Diers in \cite{Diers77}. When $\C$ has products, there is a product-preserving functor $\Pi:\;\prod(\C)\rightarrow \C$ assigning to each object $\bar{A}$ its product $\prod\bar{A}$ in $\C$. This functor describes a isomorphism between category $\C$ and $\prod(\C)$. In this case we use the product to aggregate families of objects and families of arrows. This is done using two maps $i_0:fam(\C_0)\rightarrow fam(\C_0)$ and $i_2:fam(\C_2)\rightarrow fam(\C_2)$ defined by $i_0((A_i)_I)=\prod_{i\in I}A_i$ and, when $\bar{f}=(f_j)_J:\Box \bar{f} \rightarrow \bar{f}\Box$ in $\C_2$, we define $i_2(\bar{f})=\prod_{j\in J} f_j:i_1(\Box f)\rightarrow i_1(f\Box)\in \C_1$.
Note that $i_1$ and $i_2$ are idempotents, $i_0\circ i_0 = i_0$ and $i_2\circ i_2 = i_2$, and both maps have a nonempty class of fixed points.

Given a category $\C$ with products, we have $\C\cong \prod(\C)$ then:

\begin{proposition}
Every category $\C$ with products, has the structure of multi-category.
\end{proposition}

Assuming that for every multi-morphism $f:\Box f\rightarrow f \Box$ has by dual a multi-morphism $f^{op}: f\Box\rightarrow \Box f$, is a natural consequence from definition that:
\begin{proposition}
If $\C$ is a category with structure of multi-category, then also its dual $\C^{op}$ also has the structure of multi-category.
\end{proposition}
In particular defining $\coprod(\C)$ as the dual of product completion $(\prod(\C^{op}))^{op}$, if $\C$ has coproducts then $\C\cong \coprod(\C)$ and $\C$ has the structure of multi-category.

In this sense, for the canonical embeddings $J:\C\rightarrow \prod(\C)$ and $J':\C\rightarrow \coprod(\C)$,  the multi-category $\prod(\C)$ is generated by $J(\C)$ and  the multi-category $\coprod(\C)$ is generated by $J'(\C)$.

\begin{figure}[h]
 \[
 \small
\xymatrix @=5pt {
&&&*+[o][F-]{f}\ar `r[rd][rd]\ar `r[rrd][rrd]\ar `r[rrrd][rrrd]&&&\\
 X_0\ar `u[urrr][urrr]&X_1\ar `u[urr][urr]& X_2\ar `u[ur][ur]&&X_3&X_4&X_5
 }
\]
\caption{Multi-morphism.}\label{multiarrow}
\end{figure}
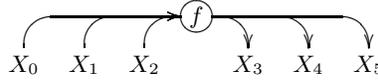

Let $\C$ be a multi-category. For every object $X$, such that $X\ncong \top$, a \emph{factorization} for $X$ is a family of objects $(X_i)_I$ such that $i_2(\cup_{I}X_i)\cong X$,  with each $X_i\ncong \top$. A morphism $f:X\rightarrow Y$ and two factorizations $i_2(\cup_{I}X_i)\cong X$ and $i_2(\cup_{J}Y_j)\cong Y$, define a \emph{multi-morphism}, denoted in this case by $f:(X_i)_I\rightarrow (Y_j)_J$, and we simplify notation by writing $\Box f= (X_i)_I$ and $f \Box =  (Y_j)_J$. In Figure \ref{multiarrow} we presented a pictographic representation for a multi-morphism $f$ with $\Box f =\{X_0,X_1,X_2\}$ and $f \Box =\{X_3,X_4,X_5\}$.

A multi-morphism can be seen as a \emph{multi-arrow} defined from a set of source nodes to a set of target nodes. A set of multi-arrows defined on a set of nodes describes a \emph{multi-graph}. In this sense a multi-graph is defined by a set of nodes and a set of multi-arrows linking a set of source nodes to a set of target nodes. Similarly to the notion of diagram:

\begin{definition}
If $\C$ has the structure of multi-category a \emph{multi-diagram} $D:\G\rightarrow \C$, is a correspondence, $D$, defined from a multi-graph $\G$ to the multi-category structure of $\C$: assigning to each node in $\G$ a object in $\C$ and to each multi-arrow a multi-morphism defined in $\C$, such that for every multi-arrow $f:\{a_1,\ldots,a_n\}\rightarrow\{b_1,\ldots,b_m\}$ in $\G$,
\[
D(f):D(a_1)\cup\ldots\cup D(a_n)\rightarrow D(b_1)\cup\ldots\cup D(b_m).
\]
\end{definition}

Following the spirited of Diers' extension from limits to multi-limits in the context of product completion \cite{Diers77}. We defined:

\begin{definition}[multi-limit]
Let $\C$ be a category and let $J:\C\rightarrow \D$ be an embedding on the multi-category $\D$ generated by $J(\C)$. Exists the multi-limit for a diagram $D:\G\rightarrow \C$, if there is an object $A$ in the multi-category $\D$ such that the multi-diagram $D\circ J$ have by limit $A$, and we write in this case $$\text{Mlim}_J D=A.$$
\end{definition}

When embedding $J:\C\rightarrow \D$ on the multi-category $\D$ generated by $J(\C)$ is described using the free product completion, $\D\cong \prod(\C)$, the above definition coincide with the Diers' extension for limits.  This type of limit extension, by the completion of the category structures, not fulfils our need of a fuzzy notion of limit. Since our goal is be able of describing a structure by approximation. For that, in the following section we describe the appropriated multi-category structure useful on a framework for description by approximation, for that we need to extend further the notion of structural completion. To the presentation of such a framework let began by presenting the multi-category of relations evaluated on a complete resituated lattice.

\begin{figure}[h]
\[
\small
\xymatrix @=7pt {
&&&&&*+[o][F-]{f}\ar `r[rdd][rdd] &\\
&&*+[o][F-]{g}\ar `r[rrd][rrd]\ar `r[rrrd][rrrd] &&&&\\
 X_0\ar `u[urr][urr]\ar `d[drrrr][drrrr]&X_1\ar `u[ur][ur]\ar `d[drrr][drrr]& &X_2\ar `u[uurr][uurr]\ar `d[dr][dr]&X_3\ar `u[uur][uur]&X_4&X_5\\
 &&&&*++[o][F-]{f\comp g}\ar `r[ru][ru]\ar `r[rru][rru]&&\\
 }
\]
\caption{A multi-diagram describing the composition of multi-morphisms.}\label{multimorphism composition}
\end{figure}
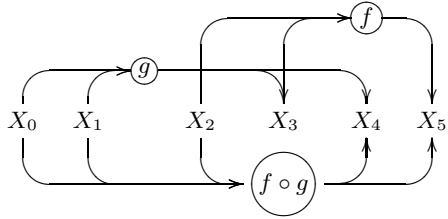

\subsection{Relations evaluated in $\Omega$}

The multi-category $Rel_\Omega$ has by objects sets of sets and by multi-morphisms relations, however these relation are evaluated in a multi-valued logic, modeled by a complete resituated lattice $\mathbf{\Omega}=(\Omega,\otimes,\Rightarrow,\wedge,\vee,\bot,\top)$. A multi-morphism $f:A\rightarrow B$ is defined by a map $f:\prod A\times \prod B\rightarrow \Omega$, and we write when this is the case $\Box f= A$ and $f \Box= B$.

A flavor for a multi-category $Rel_\Omega$ is defined by a semiring $(\Omega,\times,\top,+)$. It is defined selecting in the complete resituated lattice $\mathbf{\Omega}$, $+\in\{\oplus,\vee\}$ and $\times\in\{\otimes,\wedge\}$, such that $(\Omega,\times,\top)$ is a monoid  and $(\Omega,+)$ is a semigroup and $\times$ distributes over $+$.

Composition between multi-category in $Rel_\Omega$ is defined using the selected flavor $(\Omega,\times, \top,+)$ for the multi-category.  Given multi-morphisms $f:A\rightarrow B$ and $g:C\rightarrow D$ we define its composition in $Rel_\Omega$ by the multi-morphism $f\circ g:(\Box f \cup \Box g\backslash f \Box)\rightarrow (g \Box \cup f\Box \backslash \Box g)$ given by the map
\[
(f\circ g)(\bar{x},\bar{z})=\sum_{\bar{y}\in\prod(f\Box\cap \Box g)}f(\bar{x},\bar{y})\times g(\bar{y},\bar{z})
\]
where $\bar{x}\in\prod(\Box f \cup \Box g\backslash f \Box)$ and $\bar{z}\in\prod(g \Box \cup f\Box \backslash \Box g)$.

Note that, when $Rel_\Omega$ is governed by the classical bivalent logic, all the possible flavors coincide and the composition of composable multi-morphisms is the composition of relations. In this case, when relations are maps, the compostion of composable maps is precisely the composition of maps between sets.

The multi-graph presented in Figure \ref{multimorphism composition} describes the composition between multi-graph $g:\{X_0,X_1\}\rightarrow \{X_2,X_3,X_4\}$ and $f:\{X_2,X_3\}\rightarrow \{X_5\}$, to produce the multi-morphism $g\circ f:\{X_0,X_1, X_2\}\rightarrow \{X_4,X_5\}$.

\section{Logical extension for universal properties}

Similarity is an important concept in many research areas; for example,
in biology, computer science, linguistics, logic, mathematics, philosophy and
statistics, a great deal of work has been done on similarity issues. The main
goal of data mining is to analyze data sets and find patterns and regularities
that contain important knowledge about the data. In searching for such
regularities, it is usually not enough to consider only equality or inequality
of data objects. Instead, we need to consider how similar, or different two
objects are, i.e., we have to be able to quantify how far from each other two
objects are. This is the reason why similarity between objects
is one of the central concepts in data mining and knowledge discovery.
A notion of similarity between objects is needed in virtually any
database and knowledge discovery application.

How similarity between objects is defined, however, largely depends on
the type of the data. The objects considered in data mining are often
complex, and they are described by a different number of different kinds of
features. On the other hand, on a single set of data we
can have several kinds of similarity notions.  Different similarity measures can reflect different facets of the data, and therefore, two objects can be determined to be very similar by one measure and very different by another measure. In practice, however similarity degrees have mainly an ordinal meaning. In other words it is the ordering induced by the similarity degrees between the elements that is meaningful, rather than the exact value of the degrees. We assume similarity relations evaluated in a lattice of truth values. The same set used to describe membership grades of elements to an object, useful on the encoding of data imprecision or uncertainty. This allows the use of membership relations and similarity relations directly for predicate construction in a multi-valued logic. To this logic, used on the evaluation of all membership relations and similarity relations, we called the logic of the universe of discurse.

Despite the fact that there is no single definition for similarity, and that
one single measure seldom suits for every purpose, we try to describe
a generic framework, named $\Omega$-multi-categories, to the manipulation of objects having similarity and membership relations associated.

While our first goal for the definition of multi-categories was essentially functional, or as a framework for the relational interpretation for circuits. The idea associated with the notion of $\Omega$-multi-categories is the logical extension, the possibility of induce a multi-valued structure to those relations. In this sense we generalize the possibility of see relations, such as membership and similarity, as a matrix having by elements truth-values in $\Omega$.

In a $\Omega$-multi-categories we assume the existence of an object having by endomorphisms truth-values. This structure tries to catch the structure of $Rel_\Omega$, where for the singleton set $\{\ast\}$, $Rel_\Omega[\{\ast\},\{\ast\}]$ has by elements a endomorphism defined by each $\Omega$. Each multi-morphism in $Rel_\Omega$, $f:A\rightarrow B$ is interpreted as a matrices taken values in $\Omega$, if $(a,b)\in A\times B$, then $f(a,b)\in \Omega$ and each matric defined by a selection $f:\{\ast\}\rightarrow A\times B$, $f(a,b)=f(\ast,(a,b))$, i.e. $\D[A,B]=\D[\ast,A\cup B]$. In this sense we define:

\begin{definition}
A multi-category $\D$ is a $\Omega$-multi-category, where there is a residuated lattice $\mathbf{\Omega}=(\Omega,\otimes,\Rightarrow,\wedge,\vee,\bot,\top)$, and:
\begin{enumerate}
  \item for every pair of objects $A$ and $B$, there is a bijective map $(\_)^\circ: \D[A,B]\rightarrow \D[B,A]$ such that $(f^\circ)^\circ=f$, $(f\circ g)^\circ=g^\circ\circ f^\circ$ and $1_A^\circ=1_A$;
  \item there is an objet $\ast$ in $\D$ and a bijective map $\ulcorner\_\urcorner:\D[\ast,\ast]\rightarrow \Omega$, assigning to each endomorphism $f:\ast\rightarrow \ast$ a value $\ulcorner f \urcorner\in \Omega$, such that $\ulcorner 1_\ast\urcorner=\ast$, and for every $f,g\in \D[\ast,\ast]$, $\ulcorner f\circ g \urcorner=\ulcorner f\urcorner\otimes \ulcorner g \urcorner$ and $\ulcorner f^\circ\urcorner =\ulcorner f\urcorner$.
\end{enumerate}
In the multicategory $\D$, we also assume, for every pair of objects $A$ and $B$, $$\D[A,B]=\D[\top,A\cup B],$$
and an order $\ast$ defined in $\D[A,B]$, with top element $\ast$, and such that, given two multi-morphisms $f,g:A\rightarrow B$, with $f\leq g$, we have $g^\circ \leq f^\circ$, and given $h:C\rightarrow A$ and $i:B\rightarrow D$, $i\circ f\circ h\leq i\circ g \circ h$.
\end{definition}

\begin{example}
Every multi-category $\D$, with a terminal element $\ast$, can be extended to an $\Omega$-multi-category, if we weighted formally its morphisms using values from a resituated lattices $\Omega$. For that we begin by defining a new multi-category $\Omega(\D)$ having by objects $\D$-objects and defining formal endomorphism $\ast$ such that $\D_\Omega[\ast,\ast]=\Omega$. For every $f\in\D[A,B]$ and every $\lambda\in \Omega$ we formally define weighted multi-morphisms in $\Omega(\D)$, $(f,\lambda):A\rightarrow B$, $(f,\lambda)^\circ:B\rightarrow A$, $(f,\lambda):\top\rightarrow A\cup B$ and $(f,\lambda)^\circ:\top\rightarrow B\cup A$. The operation on multi-morphisms resultes from extending the operation in $\D$, making $$(f,\lambda_0)\comp (g,\lambda_1)=(f\comp g,\lambda_0\otimes \lambda_1).$$
The functor $J:\D\rightarrow \Omega(\D)$ such that $J(A)=A$ and $J(f)=(f,\top)$ defines a embedding.

Let $\HH$ be a $\Omega$-multi-category and every functor $F:\D\rightarrow \HH$ can be extend to multi-functor $\overline{F}:\Omega(\D)\rightarrow \HH$ between $\Omega$-multi-categories. In this sense we see $\Omega(\D)$ as a completion of $\D$ to truth-values on $\Omega$.
\end{example}

Given an $\Omega$-multi-category $\D$ and an $\D$-object $A$,  we called \emph{element} of $A$, to every $\D$-multi-morphism $a:\top \rightarrow A$.
On a category with structure of $\Omega$-multi-category we may present a satisfactory extension  to equality, using similarity relations defined between elements.

\begin{definition}
A category $\C$ has structure of $\Omega$-multi-category if it is equivalent to a $\Omega$-multi-category $\D$,
\end{definition}

\begin{definition}
Let $\D$ be a category, $\Omega$ a residuated lattice and $J:\C\rightarrow \D$ be an embedding on the $\Omega$-multi-category $\D$. $\D$ is a $\Omega$-multi-category generated by $J(\C)$ if $\D$ is a multi-morphism generated by $J(\C)$, with the structure of $\Omega$-multi-category.
\end{definition}

An $\Omega$-object in $\D$ is a triple $(A,a,\alpha)$ defined using an object $A\in\D_0$, an element $a:\top\rightarrow A\in \D_2$ and a similarity relation $\alpha:A\rightarrow A$, satisfying:
$$1_A\leq \alpha,\;\alpha = \alpha^\circ,\text{ and }\alpha \circ \alpha \leq \alpha.$$
We abbreviate the representation of a $\Omega$-object $(A,a,\alpha)$ by writing $a:\alpha$.

A morphism $f:(\alpha:A)\rightarrow (\beta:B)$ between $\Omega$-objects, is a conservative bimodule defined by a morphism $f:A\rightarrow B$ in $\D$, such what
$$f\circ a \leq b,\;f\circ a\leq f\text{ and } b\circ f\leq f.$$
The composition in $\D$ preserves this structure: for morphisms $f:(\alpha:A)\rightarrow (\beta:B)$ and $g:(\beta:B)\rightarrow (\gamma:C)$, the morphism $g\circ f$ is a morphism between $\Omega$-objects since, $g\circ f\circ a\leq g \circ b\leq c$, $g\circ f\circ \alpha \leq g \circ f$ and $\beta \circ g\circ f \leq g \circ f$.

Every $\Omega$-object $a:\alpha$ has by identity $id(A):A\rightarrow A$ in $\D$. The identity $1_\ast$, defined on object $\ast$ is a similarity relation $1_\ast\leq 1_\ast$, $1_\ast=1_\ast^\circ$ and $1_\ast\circ1_\ast\leq1_\ast$. Defining the $\Omega$-object $(\ast,\top,1_\ast)$ and every morphism $f\in \D[\ast,\ast]$ defines a multi-morphism $f:(\top:1_\ast)\rightarrow(\top:1_\ast)$.

The class of $\Omega$-objects and conservative bimodules, with the composition in $\D$, defines a category denoted by $\D_\Omega$.

\begin{lemma}
Let $R$ be a bivalente relation defined between $\Omega$-objects, in $\D_\Omega$, such that $(A,a,\alpha)R(B,b,\beta)$, if there is a morphism $f:(A,a,\alpha)\rightarrow(B,b,\beta)$ such that \[f\circ a = b\text{ and }\beta=f^\circ \circ \alpha \circ f.\]
The relation $R$ is an equivalence between $\Omega$-objects.
\end{lemma}

In $\D_\Omega$  two $\Omega$-objects $(A,a,\alpha)$ and $(B,b,\beta)$ are equivalents if there is a morphism $f:(A,a,\alpha)\rightarrow(B,b,\beta)$ such that \[f\circ f^\circ = 1_B,\;f\circ a = b\text{ and }\beta=f^\circ \circ \alpha \circ f,\]
in this case we write $(A,a,\alpha)\cong (B,b,\beta)$.

The functor $J:\D\rightarrow \D_\Omega$, given by $J(A)=(A,\top,1_A)$ and $J(f)=f$ defines an embedding.  The category $\D_\Omega$ has structure of multi-category when we define object aggregation by $i'_0(A,a,\alpha)=(\prod i_0(A),\prod i_2(a),\prod i_2(\alpha))$, and for every morphism $f:(A,a,\alpha)\rightarrow (B,b,\beta)$, the multi-arrow aggregation is given as $i'_2(f)=i_2(f)$, the aggregation in $\D$, $\Box f=(A,\top,1_A)$ and $f\Box=(B,\top,1_B)$, having by empty circuit $\top=(\top,\top,1_\top)$ and where the object class is partially ordered using relation $\subset$ such that, $(A,a,\alpha)\subset (B,b,\beta)$ if $A\subset B$, in $\C$, or ($A=B$ and $\alpha \leq \beta$) or ($A=B$ and $\alpha=\beta$ and $a\leq b$). The class of objects is algebrized with the operation $\cup$ in $\D$, $(A,a,\alpha)\cup (B,b,\beta)= (A\cup B,a\cup b,\alpha\cup \beta)$ and by $(A,a,\alpha)\backslash (B,b,\beta)= (A\backslash B,a|_{A\backslash B},\alpha|_{A\backslash B})$ defined using application restriction.

\begin{proposition}
Given a multi-category $\D$. The category $\D_\Omega$ defined by $\Omega$-objects and conservative bimodules in $\D$ has the structure of multi-category. If $\D$ is a multi-category generated by $\C$, then $\D_\Omega$ also is generated by $\C$.
\end{proposition}

Denoting by $\D_\Omega[a:\alpha,b:\beta]$ the class of every multi-morphism $f:(a:\alpha)\rightarrow(b:\beta)$, it is a subclass of $\D_\Omega[\alpha:A,\beta:B]$, the class of every multi-morphism $f:(A,a,\alpha)\rightarrow (B,b,\beta)$, we have $$\D_\Omega[a:\alpha,b:\beta]\subset \D_\Omega[\alpha:A,\beta:B]\subset \D[A,B]. $$

The logic extension equality between elements can be made for categories with the structure of $\Omega$-multi-category.

\begin{definition}
A category $\C$ has structure of $\Omega$-multi-category if it is equivalente to a $\Omega$-multi-category $\D_\Omega$ generated by $\C$. In a category with structure of $\Omega$-multi-category, two elements $a,b:\alpha$ are $\lambda$-similar, with $\lambda\in \Omega$, if $$\ulcorner b^\circ \circ \alpha \circ a\urcorner = \lambda$$
and in this case we write $[a=b]_\alpha=\lambda$.
\end{definition}

By definition in a $\Omega$-multi-category every multi-morphism $f:(a:\alpha)\rightarrow (b:\beta)$ defines an element $f:\alpha\otimes \beta$.
\[
\small
\xymatrix @=7pt {
\top \ar[rr]^{a} \ar[rrdd]_b&&A\ar[rr]^\alpha\ar[dd]_{f}&&A\ar[dd]_{f}\\
&&&&\\
&&B\ar[rr]^{\beta}&&B\\
 }
\]
\[
\small
\xymatrix @=7pt {
\top \ar[rr]^{f}&&A\cup B\ar[rr]^{\alpha\otimes \beta}&&A\cup B\\
 }
\]
Given multi-morphisms $f,g:\alpha\rightarrow \beta$  we have $[f=g]= g^\circ \circ (\alpha\otimes \beta) \circ  f$. Note what, the defined relation $[\_=\_]$ is a similarity in $\D_\Omega[\alpha:A,\beta:B]$, since $[f=f]= f^\circ \circ (\alpha\otimes \beta) \circ  f\geq  f^\circ \circ  f^\circ \geq 1_\top$, $[f=f]^\circ= f^\circ \circ (\alpha\otimes \beta) \circ  f$, $[f=g]^\circ=( g^\circ \circ (\alpha\otimes \beta) \circ f)^\circ =  f^\circ \circ (\alpha\otimes \beta) \circ  g= [g=f]$ and $[g=h]\circ [f=g] = h^\circ \circ (\alpha\otimes \beta) \circ g \circ  g^\circ \circ (\alpha\otimes \beta)\circ  f \leq  h^\circ \circ (\alpha\otimes \beta) \circ (\alpha\otimes \beta) \circ  f \leq   h^\circ \circ (\alpha\otimes \beta)^\circ \circ  f=[f=h]$.

This notion of equality is a conservative extension to the classical one. For every category $\C$ with structure of $\Omega$-multi-category, to morphisms $f,g:(a:A)\rightarrow (b:B)$ are equal in the category $\C$, $f=g$, iff $[f=g]_{1_A\otimes 1_B}=\top$ in $\D_\Omega$.

Since in a $\Omega$-multi-category every hom set $\D_\Omega[A,B]$ is partially sorted, following Freyd and Scedrov \cite{Freyd90} a morphism is:
\begin{enumerate}
  \item \emph{entire} if $1_A\leq f^\circ\circ f$, and
  \item \emph{simple} if $f\circ f^\circ\leq 1_A$.
\end{enumerate}
A morphism in $\D_\Omega$ is called a \emph{map} when it is  entire and simple. When the element $a\in \D_\Omega[\top,A]$ is defined by a map, we express this by writing $!a\in A$. Note that, for every $a\in \top$, $a:\top\rightarrow \top$ is a map and represents a truth-value in $\Omega$.

Given a morphism $f:(a:\alpha)\rightarrow(b:\beta)$ in $\D_\Omega$ we define
$$f(a,b)=b^\circ\circ f\circ a $$
a subobject of $A\cup B$. When $f(!a,!b)=\top$, we write $f(a)= b$.

When $\D$ has products, for elements $a\in A$ and $b\in B$, the unique subobject $f:\top\rightarrow A\times B$ such that, $\pi_1\circ f = a$ and $\pi_2\circ f = b$ is denoted by $a\times b \in A\times B$.

In a complete category $\D$, if $D:\G\rightarrow\D$ is a multi-diagram with vertices $(A_i)_I$, having by limit $(Lim\;D,(f_i)_I)$, it defines an element $lim\;D\in \prod_iA_i$. For that, take by $l:Lim\;D\rightarrow \prod_iA_i$ the unique morphism in $\D$ such that for every $i\in I$, $\pi_i\circ l=f_i$. Using the top element $\top\in \D[Lim\;D,\top]$, we define $lim\;D=l\circ \top$. For every $\bar{a}\in Lim\;D$, then $$l\circ \bar{a}=a_1\times a_2\times\ldots \times a_n.$$ When, for every $i\in I$, $\alpha_i$ is a similarity relation in $A_i$, the morphism $\Pi_i\alpha_i$ is a similarity in $\prod_iA_i$, and the triple $(\prod_iA_i,lim\;D, \Pi_I\alpha_i)$  defines a $\Omega$-object in $D_\Omega$.

Similarly, for every par $(R,(f_i)_I)$, where $R$ is a $\D$-object, and $f_i:R\rightarrow A_i$, for every $i\in I$, let $l:R\rightarrow \prod_iA_i$ be the unique morphism such that, for every $i\in I$, $\pi_i\circ l=\alpha_i$.  Using the top element $\top\in \D[\top,R]$, we define $F_\top(R,(f_i)_I)=l\circ \top\in \prod_iA_i$ and the triple $(\prod_iA_i,F_\top(R,(f_i)_I), \Pi_I\alpha_i)$ is a $\Omega$-object in $\D_\Omega$.
\[
\small
\xymatrix @=7pt {
 &&A_i&&\\
 &&&&\\
 R\ar[uurr]^{f_i}\ar[rr]^l&&\prod_iA_i\ar[uu]^{\pi_i}\\
 &&&&\\
 \top\ar[uurr]_{F_\top(R,(f_i)_I)}\ar[uu]_\top&&&&\\
 }
\]

\begin{definition}
Let  $\D$ be a complete category, $D:\G\rightarrow\D_\Omega$ a multi-diagram with vertices $(A_i,f_i,\alpha_i)_I$, and a pair $(R,(f_i)_I)$, with $(R,r,\alpha)$ a $\Omega$-object and multi-morphisms $f_i:(R,r,\alpha)\rightarrow (A_i,f_i,\alpha_i)$ in $\D_\Omega$. The pair $(R,(f_i)_I)$ is $\lambda$-similar to $Lim\; D$ if $$[lim\;D\;=\;F_\top(R,(f_i)_I)]_{\prod_i\alpha_i}\geq \lambda.$$
\end{definition}

In this framework the limit can be extended to $\lambda$-limit. Note what, every element $a\in \prod_{i\in I}A_i$ can be extended to  $\bar{a}\in \prod_{j\in J}A_j$, with $(A_i)_I\subset (A_j)_J$, given by $\bar{a}=\pi^\circ \circ a$, where $\pi: \prod_{j\in J}A_j\rightarrow \prod_{i\in I}A_i$ is the projection. This simplifies the use and the manipulation of multi-morphism, and we called it the \emph{canonical extension} of $a$ to $\prod_{j\in J}A_j$.

\begin{definition}\label{multilimit}
If $\C$ is a category and $D:\G\rightarrow\D$ is a multi-diagram in $\D$ with vertices $(A_i)_I$. Let $\D$ be a complete multi-category generated by $\C$.  The weighted limit of $D$, on $\D_\Omega$  is computed assigning weights and a similarity to each $A_i$, defining $\Omega$-objects $(A_i,a_i,\alpha_i)$, compatible with the diagram structure. By this we mean what this selection must define a multi-diagram $\bar{D}:\G\rightarrow\D_\Omega$, where $\bar{D}(A_i)=(A_i,a_i,\alpha_i)$ and every multi-morphism $D(f):A_i\rightarrow A_j$ is assigned to a  conservative bimodule $\bar{D}(f):(A_i,a_i,\alpha_i)\rightarrow (A_j,a_j,\alpha_j)$ in $\D_\Omega$. The weighted limit of $D:\G\rightarrow\C$, is the element $(lim\;D):\top\rightarrow\prod_{i\in I}A_i$ given, for the flavor $(\Omega,\times,1,+)$ in $\D_\Omega$,  by
\[
(Mlim\;D)(\bar{x})=\prod_{f\in \G}\overline{D(f)}(\bar{x}),
\]
where $\overline{D(f)}$ is the canonical extension of $D(f)$ to $\prod_{i\in I}A_i$.
\end{definition}

This notion can be applied to every diagram $\bar{D}:\G\rightarrow\D_\Omega$, but its use on extension of limits dependes on the appropriated selection of $\Omega$-objects.

\begin{definition}
Given a multi-diagram $\bar{D}:\G\rightarrow\D_\Omega$ with vertices $(A_i,f_i,\alpha_i)_I$, and a pair $(R,(f_i)_I)$, with $(R,r,\prod_i\alpha_i)$ a $\Omega$-object and multi-morphisms $f_i:(R,r,\alpha)\rightarrow (A_i,f_i,\alpha_i)$ in $\D_\Omega$, is a $\lambda$-limit of $\overline{D}$ if $$[Mlim\;\overline{D}\;=\;F_\top(R,(f_i)_I)]_{\prod_i\alpha_i}\geq \lambda.$$
\end{definition}

In the following is presented an example of a $\Omega$-multi-category generated from the category $Rel_\Omega$ of sets and relations, evaluated on a multi-valued logic $\Omega$, and flavor $(\Omega,\times,\top,+)$.

\section{The $\Omega$-multi-category $Rel_\Omega$}

Given an CRlattice $\mathbf{\Omega}=(\Omega,\otimes,\Rightarrow,\wedge,\vee,\bot,\top)$, where we select a flavor given by a semiring $(\Omega,\times,\top,+)$ used on relation composition.

A $\Omega$\emph{-set} is a triple $(A,a,\alpha)$, denoted as $a:\alpha$, with $A$ a set, $a:A\rightarrow \Omega$ a distribution defined by a map and $\alpha:A\times A\rightarrow \Omega$ a similarity relation evaluated in $\Omega$, such that $\alpha \circ a\leq \alpha$.

If $A=\{(A_i,a_i,\alpha_i)\}_I$ and $B=\{(B_j,b_j,\beta_j)\}_J$ are sets of $\Omega$-sets then $f:A\rightarrow B$ is a multi-morphism if it is a map $f:\prod_IA_i\times \prod_JB_j\rightarrow\Omega$ such that $f(\bar{x},\bar{y})\times(\Pi_Ia_i)(\bar{x})\leq (\Pi_Jb_j)(\bar{y})$, $f(\bar{x},\bar{y})\times(\Pi_I\alpha_i)(\bar{x})\leq f(\bar{x},\bar{y})$ and $(\Pi_J\beta_j)(\bar{y})\times f(\bar{x},\bar{y}) \leq f(\bar{x},\bar{y})$. When this is the case we write $\Box f =\{(A_i,a_i,\alpha_i)\}_I$ and $f\Box  =\{(B_j,b_j,\beta_j)\}_J$.
\[
\small
\xymatrix @=7pt {
\top \ar[rr]^{\Pi_Ia_i} \ar[rrdd]_{\Pi_Jb_j}&&\prod_IA_i\ar[rr]^{\prod_I\alpha_i}\ar[dd]_f&&\prod_IA_i\ar[dd]_f\\
&&&&\\
&&\prod_JB_j\ar[rr]^{\prod_J\beta_j}&&\prod_JB_j\\
 }
\]

In a $\Omega$-set $(A,a,\alpha)$, when $a$ and $\alpha$ are bivalent evaluations, $a$ describes a subset of $A$ and $\alpha$ is an equivalent relation. The top element $\top:\top\rightarrow A$, is defined by $\top(a)=\top$, for every $a\in A$ and describes $A$.

Given $\Omega$-sets $a:\alpha$ and $b:\beta$, a $\Omega$\emph{-map} $f:(a:\alpha)\rightarrow (b:\beta)$ is a map $f:A\rightarrow B$ such that, for each $x,y\in A$, $a(x)\leq b(f(x))$ and $\alpha(x,y)\leq \beta(f(x),f(y))$.

If $a:\top\rightarrow A$ is a map between sets, $a$ describes the selection of an element in $A$, and we write $!a:\top\rightarrow A$ or $!a\in A$. By $!a\in A\times B$ we define the selection of a pair $(x,y)\in A\times B$.

Composition between $\Omega$-sets is defined using a flavor described by a semiring $(\Omega,\times, \top, +)$. We define
\[
(f\circ g)(!\bar{x},!\bar{z})=\sum_{!\bar{y}\in\prod(f\Box\cap \Box g)}f(!\bar{x},!\bar{y})\times g(!\bar{y},!\bar{z})
\]
where $!\bar{x}\in\prod(\Box f \cup \Box g\backslash f \Box)$ and $!\bar{z}\in\prod(g \Box \cup f\Box \backslash \Box g)$.

Independent from the  $Rel_\Omega$ flavor, for every $\Omega$-set $(A,a,\alpha)$, its identity is the identity map $1_A:A\rightarrow A$ between sets.
\[
\small
\xymatrix @=7pt {
\top \ar[rr]^{a} \ar[rrdd]_a&&A\ar[rr]^\alpha\ar[dd]_{1_A}&&A\ar[dd]_{1_A}\\
&&&&\\
&&A\ar[rr]^{a}&&A\\
 }
\]
A multi-morphism $f:(A,a,\alpha)\rightarrow (B,b,\beta)$ is \emph{bivalente} if $f(!x,!y)\in\{\bot,\top\}$. In this sense every identity of $\Omega$-sets is bivalente, moreover:

\begin{theorem}
In $Rel_\Omega$ independently of its flavors the composition of componible bivalente multi-morphisms is a bivalente multi-morphism. In particular, the composition of maps is a map.
\end{theorem}

Given a CRlattice, $\mathbf{\Omega}$, and an associated semiring $(\Omega,\times,\top,+)$. The set of $\Omega$-sets, multi-morphism and the composition with flavor $(\Omega,\times,\top,+)$ define a multi-category, denoted by $Rel_\Omega$. Where object aggregation is given by the cartesian product
\[
i_0(\{(A_i,a_i,\alpha_i)\}_I)=\{(\prod_I A_i,\prod_Ia_i,\prod_I\alpha_i)\}
\]
and multi-morphism interpretation is defined for $f:A\rightarrow B$, with $A=\{(A_i,a_i,\alpha_i)\}_I$ and $B=\{(B_j,b_j,\beta_j)\}_J$, by
\[i_2(f):\{(\prod_I A_i,\prod_Ia_i,\prod_I\alpha_i)\}\rightarrow \{(\prod_J B_j,\prod_Jb_j,\prod_J\beta_j)\}\]
given by $i_2(f):\prod_IA_i\times \prod_JB_j\rightarrow\Omega$, $i_2(f)(\bar{x},\bar{y})=f(\bar{x},\bar{y})$. We denoted by $\top$ the $\Omega$-set $\{(\{\ast\},\top,1_\ast)\}$ defined using the singleton set $\top=\{\ast \}$ and its identity map $1_\top:\top \rightarrow \top$.

The class of sets of $\Omega$-sets has a monoidal structure defined by set union and having by identity the empty set. And its is partially sorted by set inclusion.

Two sets of $\Omega$-sets $A=\{(A_i,a_i,\alpha_i)\}_I$ and $B=\{(B_j,b_j,\beta_j)\}_J$ are equivalent if exists a multi-morphism $f:A\rightarrow B$ such that $f\circ \Pi_Ia_i=\Pi_Jb_j$ and $f^\circ\circ(\prod_I\alpha_i)\circ f = \beta$, in this case we write $\{(A_i,a_i,\alpha_i)\}_I\cong \{(B_j,b_j,\beta_j)\}_J$.

Since $f\in Rel_\Omega[A,B]$ is by definition a map $f:\prod A\times \prod B\rightarrow \Omega$ and $\top\times \prod A\times \prod B \cong \prod A\times \prod B$, we have $Rel_\Omega[\top,A\cup B]\cong Rel_\Omega[A, B]$. And each homset $Rel_\Omega[A, B]$ has a natural order, defined extending the order in $\Omega$. For every $f,g\in Rel_\Omega[A, B]$ we have
\[
f\leq g\text{ iff } f(!\bar{x},!\bar{y})\leq g(!\bar{x},!\bar{y}),\text{ for every } !\bar{x}\in A,!\bar{y}\in B.
\]
For every $A$ and $B$ the top element $\top$ in $Rel_\Omega[A,B]$ is $\top(!\bar{x},!\bar{y})=\top$, and for $f\in Rel_\Omega[A,B]$, $f^\circ(!\bar{y},!\bar{x})=f(!\bar{x},!\bar{y})$, defines an isomorphism between $Rel_\Omega[A,B]$ and $Rel_\Omega[B,A]$.

A multi-morphism $f:(A,a,\alpha)\rightarrow (B,b,\beta)$ is a map in $Rel_\Omega$ if $f:A\rightarrow B$ is a map in $Set$. The functor $J:Set\rightarrow Rel_\Omega$ such that $J(A)=(A,\top,1_A)$ and $J(f)=f$ is an embedding.

Every diagram $D:\G \rightarrow Set$, with vertices $(A_i)_I$, defines a multi-diagram in $Rel_\Omega$, using the embedding $J:Set\rightarrow Rel_\Omega$, and defined by $J\circ D:\G \rightarrow Rel_\Omega$, having by vertices $(A_i,\top,1_{A_i})_I$. The limit of $D$ in $Set$, $(Lim\;D,(\alpha_i)_I)$ defines a cone in $Rel_\Omega$, having by vertex $(Lim\;D,\top,1_{Lim\;D})$ where by definition $Lim\;D$ is a subset of $\prod_IA_i$.
\[
\small
\xymatrix @=7pt {
\top \ar[rr]^\top \ar[rrdd]_\top&&Lim\;D\ar[rr]^{1}\ar[dd]_{J(!f)}&&Lim\;D\ar[dd]_{J(!f)}\\
&&&&\\
&&\prod_IA_i\ar[rr]^{1}&&\prod_IA_i\\
 }
\]
The relation $J(!f)\circ \top: \top \rightarrow \prod_IA_i$ defines an element in $\prod_IA_i$, denoted by $lim\;D$, and we write $lim\;D\in\prod_IA_i$. For every element $x\in \prod_IA_i$, its similarity with the limit $lim\;D\in\prod_IA_i$ is given when we fixed similarity relations $\alpha_i$ to each $A_i$, and it is given by
\[
[x = lim\; D]_{\Pi_I\alpha_i}=x^\circ\circ\Pi_I\alpha_i\circ lim\;D\vee (lim\;D)^\circ\circ\Pi_I\alpha_i\circ x.
\]
When $[x = lim\; D]_{\Pi_I\alpha_i}\geq \lambda$, we call the element $x:\top\rightarrow \prod_IA_i$ a quasi $\lambda$ limit for $D$.

In the framework $Rel_\Omega$ we extend the notion of limit in $Set$, according with Definition \ref{multilimit}, for every multi-diagram $D:\G\rightarrow Rel_\Omega$ with vertices $(A_i,a_i,\alpha_i)_I$
by
\[
(Mlim\;D)(\bar{x})=\prod_{f\in \G}\overline{D(f)}(\bar{x}),
\]
where $\overline{D(f)}$ is the canonical extension of $D(f)$ to $\prod_{i\in I}A_i$.

A relation $x:\top \rightarrow \prod_{i\in I}A_i$ is called the $\lambda$-limite of multi-diagram $D:\G\rightarrow Set_\Omega$ if $[x = lim\; D]_{\Pi_I\alpha_i}\geq \lambda$.

\begin{theorem}
If $D:\G\rightarrow Set$ is a diagram in the category $Set$, with vertices $(A_i)_I$, then the canonical embedding defines a multi-diagram $J\circ D:\G\rightarrow Set_\Omega$ and its multi-limit is equal to its limit element, i.e.
\[
[lim\;J\circ D = Mlim\;J\circ D]_1=\top.
\]
\end{theorem}

In this sense we see multi-limit as a conservative extension to the notion of limit in $Set$.

If the limit is a subset of $\prod_I A_i$, in $Set$, and it was extended to the definition of fuzzy elements of $\prod_I A_i$, in $Set_\Omega$. By definition a colimit can be used, in $Set$, for describing equivalence relations in $\coprod_I A_i$, and we can use them on the definition of similarity relations in $\coprod_I A_i$, in $Set_\Omega$.

Let $(coLim\;D, (l_i)_I)$ be a colimit cocone for diagram $D:\G\rightarrow Set$. In $Rel_\Omega$ this describes a multi-diagram $J\circ D:\G\rightarrow Rel_\Omega$ and, by definition of colimit in $Set$, there is a equivalence relation $\alpha$ in $\coprod_I A_i$ such that
\[
(\coprod_I A_i,\top, \alpha)\cong (colim\;D,\top,1).
\]
By definition of equivalence in $Rel_\Omega$, this can be expressed by the following diagram
\[
\small
\xymatrix @=7pt {
\top \ar[rr]^\top \ar[rrdd]_\top&&\coprod_IA_i\ar[rr]^{\alpha}\ar[dd]_{J(!f)}&&coprod_IA_i\ar[dd]_{J(!f)}\\
&&&&\\
&&coLim\;D\ar[rr]^{1}&&coLim\;D\\
 }
\]
by $1=J(!f)^\circ \circ \alpha \circ J(!f)$...

Note that, every similarity relation, is a multi-morphism. If $\alpha$ is a similarity in $A$, then it is a multi-morphism $\alpha:(A,\top,1_A)\rightarrow (A,\top,1_A)$, defining an element $\alpha:\top\rightarrow A\times A$. In this sense an element $\alpha:\top\rightarrow \coprod_IA_i\times \coprod_IA_i$ is a $\lambda$-approximation to the colimit of $D:\G\rightarrow Set$ in $Set_\Omega$ with vertices assignments $(A_i,a_i,\alpha_i)_I$ if
$$[x = colim\; D]_{\Pi_I\alpha_i\times \Pi_I\alpha_i}\geq \lambda.$$

\section{Weighted limits in $Rel_{\Omega}$}\label{multidiagrams}

The categorical notion of limits in $Set$ can be described using diagram tabulation\footnote{Also called diagram internalization}\cite{Borceux94}. The limit for a diagram  $D:\G\rightarrow Set$, with vertices $V=\{X_i\}_{i\in I}$ and arrows $A=\{f_j\}_{j\in J}$, is a table or a subset of $\prod_{i\in I}D(X_i)$ given by
\begin{equation}
Lim\;D=\{(\ldots,x_i,\ldots,x_j,\ldots)\in \prod_{i}D(X_i):\forall_{f:X_i\rightarrow X_j}D(f)(x_i)=x_j\}.
\end{equation}
This is a central concept on the description of algebraic structures, however for our intentions the notion is very ``crisp''.

We present here a soft conservative extension of this notion in $Rel_{\Omega}$, for a flavor $(\Omega,\times,\top,+)$. For a multi-diagram like the one presented in Figure \ref{multidiagram1}, we take as its multi-limit an $\Omega$-relation
\[
Lim\;D:X_0\times X_1\times X_2\times X_3\times X_4\times X_5\rightarrow \Omega,
\]
 such that $(Lim\;D)(\bar{x})\subset \prod_iA_i$, for every $\bar{x}\in X_0\times X_1\times \cdots\times X_5$. More precisely, for each $\bar{x}=(x_0,x_1,x_3,x_4,x_5)\in X_0\times X_1\times \cdots\times X_5$, we define
\begin{center}
\[
(Lim\;D)(\bar{x})=f(x_0,x_1,x_3,x_4,x_5)\times g(x_1,x_2,x_4,x_5)\times h(x_2,x_3).
\]
\end{center}
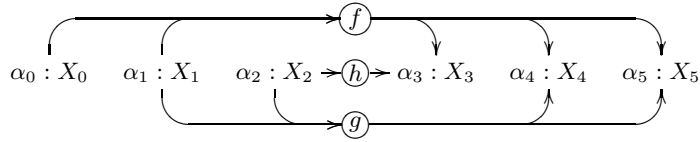
\begin{figure}[h]
\[
\small
\xymatrix @=7pt {
&&&*+[o][F-]{f}\ar `r[rd][rd]\ar `r[rrd][rrd]\ar `r[rrrd][rrrd]&&&\\
 \alpha_0:X_0\ar `u[urrr][urrr]&\alpha_1:X_1\ar `u[urr][urr]\ar `d[drr][drr]& \alpha_2:X_2\ar `d[dr][dr]\ar[r]&*+[o][F-]{h}\ar[r] &\alpha_3:X_3&\alpha_4:X_4&\alpha_5:X_5\\
 &&&*+[o][F-]{g}\ar `r[rru][rru]\ar `r[rrru][rrru]&&&
 }
\]
\caption{A multi-diagram $D:\G\rightarrow Set_\Omega$.}\label{multidiagram1}
\end{figure}

Bellow we present some examples of weighted limits.

\begin{example}[Weighted equalizers]
A diagram $D$ defined using two parallel morphisms $f,g:(a:\alpha)\rightarrow (b:\beta)$, has by limit a relation having by support $X\times Y$, and given by
\begin{equation}
(Lim\;D)(x,y)= f(x,y)\times g(x,y).
\end{equation}
\end{example}

\begin{example}[Weighted pullback]
A diagram  $D$ is $Set_\Omega$ defined by $f:(a:\alpha)\rightarrow (c:\gamma)$, and $g:(b:\beta)\rightarrow (c:\gamma)$,
has by limit a relation with support $X\times Y\times Z$ given by
\begin{equation}
(Lim\;D)(x,y,z)= f(x,z)\times g(y,z).
\end{equation}
\end{example}

\begin{example}\label{example1}
Let $f:A\times B\rightarrow C, g:A\times B\rightarrow C\times D\text{ and } h:A\times C\rightarrow E,$ be morphisms, with supports describe by $\Omega$-sets $\alpha:A, \beta:B, \gamma:C, \delta:D\text{ e } \epsilon:E,$ where $A=B=C=D=\{0,1\}$ and $\alpha=\beta=\gamma=\delta=\epsilon=1_{\{0,1\}\times\{0,1\}}$ the identity relation. If we assume each morphism described by the tables bellow defined in $Rel_{[0,1]}$  governed by product logic (in this tables the missing cases are supposed to have weighted zero).
\begin{center}
\begin{tabular}{ccc}

$f:\;$\begin{tabular}{|c|c||c||c|}
  \hline
  A & B & C & $\Omega$ \\
  \hline
  1 & 0 & 0 & 1 \\
  0 & 1 & 0 & 1/2 \\
  1 & 1 & 0 & 1/2 \\
  0 & 0 & 0 & 1 \\
  0 & 0 & 1 & 1 \\
  1 & 1 & 1 & 1 \\
  \hline
\end{tabular}&

$g:\;$\begin{tabular}{|c|c||c|c||c|}
  \hline
  A & B & C & D & $\Omega$ \\
  \hline
  0 & 1 & 0 & 1 & 1 \\
  1 & 1 & 0 & 1 & 1/2 \\
  0 & 0 & 0 & 1 & 1 \\
  1 & 1 & 1 & 0 & 1/2 \\
  \hline
\end{tabular}&

$h:\;$\begin{tabular}{|c|c||c||c|}
  \hline
  A & C & E & $\Omega$ \\
  \hline
  1 & 1 & 1 & 1/2 \\
  0 & 0 & 1 & 1 \\
  1 & 0 & 1 & 1/2 \\
  \hline
\end{tabular}
\end{tabular}
\end{center}
Morphisms $f,g,h$ define a diagram $D$, in $Rel_{[0,1]}$, having by limit a relation $Lim\;D$ with support $A\times B\times C\times D$ such that, for $\bar{x}=(1,0,0,0,1)$ and $\bar{x}=(1,1,0,1,1)$, we have respectively,
$Lim\;D(1,0,0,0,1)= f(1,0,0)\times g(1,0,0,0)\times h(1,0,1)=1\times 1\times 0\times 1/2 = 0,$ and
$Lim\;D(1,1,0,1,1)= f(1,1,0)\times g(1,1,0,1)\times h(1,0,1)=1\times 1/2\times 1/2\times 1/2 = 1/8$. The resulting table is presented bellow.
\begin{center}
\begin{tabular}{c}
$Lim\;D:\;$
\begin{tabular}{|c|c|c|c|c||c|}
  \hline
  A & B & C & D & E & $\Omega$ \\
  \hline
  0 & 1 & 0 & 1 & 1 & $1/2\times 1 \times 1= 1/2$ \\
  1 & 1 & 0 & 1 & 1 & $1/2\times 1/2 \times 1/2=1/8$ \\
  0 & 0 & 0 & 1 & 1 & $1\times 1 \times 1=1$ \\
  1 & 1 & 1 & 0 & 1 & $1\times 1/2 \times 1/2=1/4$ \\
  \hline
\end{tabular}\\
\end{tabular}
\end{center}

\end{example}

It is an immediate consequence of the definition:
\begin{proposition}[Existence of weighted limits in $Set_\Omega$]
Every multi-diagram $D:\G\rightarrow Rel_{\Omega},$ has weighted limits. When $V$ is a set of vertices of $\G$, there is a $\Omega$-relation $f\leq \prod_{X_i\in V}D(X_i)$, such what ${Lim\;D=f}$.
\end{proposition}

%---------
\subsection{Commutative multi-diagrams}\label{def:diagComu}
The commutativity of a diagram $D:\G\rightarrow Set$  can be detected in its tabular internalization $Lim\;D$. The commutativity of the diagram
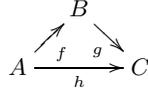
\[
\small
\xymatrix @=10pt {
&B\ar[dr]_g&\\
A\ar[ur]_f\ar[rr]_h&&C\\
}
\]
can be expressed by the equality  $f\circ g=h,$ and it is true if and only if, for every $a\in A$, we have
\begin{equation}
\bigvee_{b\in B, c\in C}f(a,b)\times g(b,c)\times h(a,c)=\top.
\end{equation}
In this sense, $A$ is called \emph{the diagram source} in $D$.

For our fuzzy conservative extension to the notion of diagram commutativity, of a multi-diagram $D$ in $Rel_\Omega$,  we began by selecting a set  $s(D)$ of its vertices. Vertices in $s(D)$ are called the \emph{diagram sources}. The commutativity is defined as a relation on those vertices.

\begin{definition}[Diagram commutativity]\label{Comutatividade diagramas}
 Let $D$ be a graph with vertices $V=\{\alpha_i:X_i\}_{i\in I}$, and let $s(D)\subset V$ be a subset of \emph{source vertices}. Assuming that $D$ doesn't have cycles involving vertices on $s(D)$, and
$
P=\prod_{X_i\in V\setminus s(D)}X_i,
$
 the cartesian product for vertices in $D$ but not in $s(D)$. The diagram $D$ is commutative in $s(D)$ iff
\begin{equation}
 \bigvee_{\bar{n} \in P}(Lim\;D)(\bar{s}\cup\bar{n})=\bigvee_{\bar{n} \in P}[\bar{s}\cup\bar{n}]_{\bigotimes \alpha_i},
\end{equation}
for every $\bar{s}\in \prod_{X_i\in s(D)}X_i$, i.e. (in the boolean case) every vector $\bar{s}$, having by componentes entities on the source vertices, can be completed with a vector $\bar{n}$ in $P$ such that, the completion is in  $Lim\;D$.
\end{definition}

\begin{example}
From Example \ref{example1}, we have $\bigvee_{b,c,d,e}(Lim\;D)(0,b,c,d,e)=1$ and $\bigvee_{b,c,d,e}(Lim\;D)(1,b,c,d,e)=1/4$. Because $\alpha=\beta=\gamma=\delta=\epsilon=1_{\{0,1\}\times\{0,1\}}$ are the identity relation $\bigvee_{b,c,d,e}(\alpha\times\beta\times\gamma\times\delta\times\epsilon)(a,b,c,d,e)=1$, for every $a\in\{0,1\}$. The multi-diagram is non-commutativity in $\{A\}$.
\end{example}

\begin{example} Let $Set_{[0,1]}$ be governed by the product logic, $\mathds{R}$ be the set of real numbers and let $+$ be the $[0,1]$-relation $+:\mathds{R}\times\mathds{R}\nrightarrow \mathds{R}$ described using the gaussian   $+(x,y,z)=e^{-\frac{(z-x-y)^2}{2}}.$ In the multi-diagram $D$, presented on Figure \ref{equation1}, each vertices is interpreted as a $[0,1]$-relations $\alpha_0:\mathds{R}$, $\alpha_1:\mathds{R}$ and $=:\mathds{R}$,
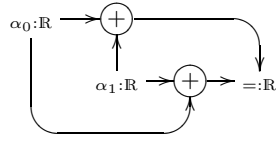
\begin{figure}[h]
\[
\small
\xymatrix @=10pt {
&&&&\\
 _{\alpha_0:\mathds{R}}\ar[r]\ar `d[ddrr]`r[rru][drr] &*+[o][F-]{+}\ar `r[rrd][rrd] &&&\\
&_{\alpha_1:\mathds{R}}\ar[u]\ar[r] &*+[o][F-]{+} \ar [r]&_{=:\mathds{R}}\\
&&&&\\
}
\]
\caption{A multi-diagram encoding $x+y=y+x$.}\label{equation1}
\end{figure}
given by $\alpha_0(x,y):=e^{-\frac{(x-x_0)^2}{2}-\frac{(y-x_0)^2}{2}},$
$\alpha_1(x,y):=e^{-\frac{(x-x_1)^2}{2}-\frac{(y-x_1)^2}{2}}$ and
$=(x,y):=\left\{
                \begin{array}{cc}
                    1 &\text{se } x=y \\
                    0 & \text{se } x\neq y \\
                  \end{array}
                \right.$,
where $x_0$ and $x_1$ are generic real parameters. The relation $+$ defines a bimodule $+:(\alpha_0\otimes\alpha_1:\mathds{R}^2)\rightarrow\; (=:\mathds{R})$.
Using the notion of weighted limit presented on Definition \ref{limit}, for every $x,y,w\in\mathds{R}$ we have,
\[
\begin{array}{rcl}
  (Lim\;D)(x,y,w) & = & +(x,y,w)\otimes+(y,x,w)\otimes[x,y,w]\\
                  & = & e^{-\frac{(w-x-y)^2}{2}}.e^{-\frac{(w-y-x)^2}{2}}.e^{-\frac{(x-x_0)^2}{2}-\frac{(x-x_0)^2}{2}}.e^{-\frac{(y-x_1)^2}{2}-\frac{(y-x_1)^2}{2}}.1 \\
                  & = & e^{-(w-x-y)^2-(x-x_0)^2-(y-x_1)^2}
\end{array}
\]
since, $e^{-(w-x-y)^2}\leq 1$, we have $e^{-(w-x-y)^2-(x-x_0)^2-(y-x_1)^2}\leq e^{-(x-x_0)^2-(y-x_1)^2}$, and
\[
\begin{array}{rcl}
  \bigvee_w(Lim\;D)(x,y,w) & = & e^{-(x-x_0)^2-(y-x_1)^2} \\
                           & = & [x]_{\alpha_0}\otimes[y]_{\alpha_1} \\
                           & = & \bigvee_w(\alpha_0\times\alpha_1\times\alpha_2)(x,y,w).
\end{array}
\]
Since  $\bigvee_w(Lim\;D)(x,y,w)=\bigvee_w(\alpha_0\times\alpha_1\times\alpha_2)(x,y,w)$, for every $x,y\in\mathds{R}$, the diagram presented in Figure \ref{equation1} is commutative.
\end{example}

\begin{example}
From Example \ref{example1}, we have $\bigvee_{b,c,d,e}(Lim\;D)(0,b,c,d,e)=1$ and $\bigvee_{b,c,d,e}(Lim\;D)(1,b,c,d,e)=1/4$. Since $\alpha=\beta=\gamma=\delta=\epsilon=1_{\{0,1\}\times\{0,1\}}$ are the identity relation, for every $a\in\{0,1\}$, $\bigvee_{b,c,d,e}(\alpha\times\beta\times\gamma\times\delta\times\epsilon)(a,b,c,d,e)=1$. Then, on product logic, because
\[
\bigwedge_a(\bigvee_{b,c,d,e}(Lim\;D)(a,b,c,d,e) \leftrightarrow \bigvee_{b,c,d,e}(\alpha\times\beta\times\gamma\times\delta\times\epsilon)(a,b,c,d,e))=
\]
\[
=(1/4 \leftrightarrow 1)\wedge (1 \leftrightarrow 1)=1/4,
\]
 multi-diagram $D$ is $1/4$-almost commutative in $\{A\}$.
\end{example}

\section{Conclusions and future work}
We are working on a framework to specify fuzzy structures, described using limits and the commutativity of multi-diagrams, able to be used for data modeling. We are investigating the possibility of this specifications be enriched, using insights extracted from data, applying machine learning tools \cite{Leandro09}.

\bibliographystyle{unsrt}
\bibliography{tesebib2}
\end{document}